\documentclass[12pt]{iopart}
\def\be{\begin{equation}}
\def\ee{\end{equation}}

\def\Zop{\bbbz}

\def\I{{\cal I}}
\def\half {\frac{1}{2}}
\def\bbbz {{\sf Z\!\!Z}}

\newcommand{\ket}[1]{|#1\rangle}

\usepackage{iopams}  
\begin{document}
% Journal identifier can be put here if required, e.g.
\jl{6}

\begin{flushright}
  hep-th/9908126\\
  CALT-68-2238\\
  DAMTP-1999-110
\end{flushright}
\vskip 0.3cm

\title[Non-BPS Dirichlet branes]{Non-BPS Dirichlet
branes\footnote[3]{Talk presented by M.R.G. at Strings '99, Potsdam,
July 19-24 1999.}}

\author{Oren Bergman\dag\ and Matthias R Gaberdiel\ddag}

\address{\dag\ California Institute of
Technology, Pasadena, CA 91125, USA.}

\address{\ddag\ Department of Applied Mathematics and Theoretical
Physics, University of Cambridge, Silver Street, Cambridge CB3 9EW,
U.K.}

\begin{abstract}
A brief introduction to the boundary state approach to Dirichlet
branes is given. The example of the non-BPS D-string of Type IIA on K3
is analysed in some detail, and its dual heterotic state is
identified.  
\end{abstract}


\submitted

\maketitle

\section{Introduction}
\setcounter{equation}{0} 

In the past year, non-BPS Dirichlet branes (D-branes) have attracted a
fair amount of attention \cite{SenRev,LR}. There exist in essence two
different approaches to constructing and analysing these states. In
one approach that has been pioneered by Sen
\cite{Sen1,Sen2,Sen3,Sen4,Sen5} (see also Sen's talk at this
conference), the non-BPS D-branes are constructed as bound states of
brane-anti-brane pairs. This construction has been interpreted in
terms of K-theory by Witten \cite{Witten2}, and this has opened the
way for a more mathematical treatment of D-branes
\cite{Horava,Gukov,BGH}. In the other approach, D-branes (and in
particular non-BPS D-branes) are described as coherent states in the
closed string theory that satisfy a number of consistency conditions 
\cite{PolCai,CLNY,Li,GrGut,BG1,BG2}. This approach will be explained 
in more detail in the next section.

The motivation for studying non-BPS D-branes is three-fold.
First, in order to understand the strong/weak coupling dualities
of supersymmetric string theories in more detail, it is important to
analyse how these dualities act on states that are not BPS
saturated. Since the dualities typically map perturbative states to
non-perturbative (D-brane type) states, we will naturally encounter
non-BPS D-branes in these considerations. Similarly, if we are to make
progress in analysing the possible dualities of theories without
supersymmetry\footnote{Some progress in this direction has recently
been made in \cite{BG1,BD,BG4,BK}.}, we have to develop
techniques to treat Dirichlet branes without supersymmetry. Second,
since the field theories describing the low energy dynamics of non-BPS
D-branes are non-supersymmetric gauge theories, configurations of
non-BPS D-branes might be useful for studying non-perturbative 
aspects of the corresponding field theories. Finally, non-BPS D-branes
offer the intriguing possibility of string compacitifications in which
supersymmetry is preserved in the bulk but broken on the brane.
\smallskip

This talk is based primarily on \cite{BG3}; more details about the
boundary state approach can also be found in 
\cite{BG1,BG4}.

\section{Dirichlet branes from boundary states}

A D-brane can be represented by a coherent (boundary) state of the 
closed string theory that describes the D-brane as a source for 
closed strings. 
Such a state describes a submanifold on which open 
strings can end provided that it satisfies
\be
\begin{array}{lclr}
(\partial X^i - \bar\partial X^i ) |B\rangle & = & 0 \qquad
\hbox{$i\;$ Neumann direction,} \\
(\partial X^I + \bar\partial X^I ) |B\rangle & = & 0 \qquad
\hbox{$I\;$ Dirichlet direction,} 
\end{array}
\ee
where $X^\mu$ denotes the coordinate field.
In the following we shall always work in the light-cone NS-R formalism.
The directions $\mu=0,1$ will be taken to be Dirichlet, so the states
are actually related by a double Wick rotation to normal D-branes
\cite{GrGut}.     

For a D$p$-brane, the above conditions can be rewritten in terms of
modes as  
\be
\label{Dirichlet}
\left.
\begin{array}{lcl}
(\alpha^\mu_n - \widetilde\alpha^\mu_{-n}) |Bp,\eta\rangle & = & 0 \\
(\psi^\mu_r - i \eta \widetilde\psi^\mu_{-r}) |Bp,\eta\rangle & = & 0
\end{array}
\right\} \mu=2,\ldots, p+2 
\ee
\be
\label{Neumann}
\left.
\begin{array}{lcl}
(\alpha^\mu_n + \widetilde\alpha^\mu_{-n}) |Bp,\eta\rangle & = & 0 \\
(\psi^\mu_r + i \eta \widetilde\psi^\mu_{-r}) |Bp,\eta\rangle & = & 0
\end{array}
\right\} \mu=p+3,\ldots, 9\,,
\ee
where $\alpha^\mu_n,\widetilde\alpha^\mu_n$ and 
$\psi^\mu_r,\widetilde\psi^\mu_r$ are the left and right-moving modes 
of the bosonic and fermionic fields, respectively, and $\eta=\pm$
describes the two different spin structures. For each choice of $\eta$
and in each left-right-symmetric sector of the theory ({\it i.e.} in
the NS-NS and the R-R sectors), a unique solution (up to
normalisation) to these equations exists  
\begin{eqnarray}
|Bp,\eta\rangle & = & \exp\left\{ \sum_{n>0} 
\left[ -{1\over n} \sum_{\mu=2}^{p+2} 
	\alpha^\mu_{-n} \widetilde\alpha^\mu_{-n} 
+ {1\over n} \sum_{\mu=p+3}^{9} 
	\alpha^\mu_{-n} \widetilde\alpha^\mu_{-n} \right] \right.
\nonumber \\
& & \qquad \qquad \left. + i \eta \sum_{r>0}
\left[ - \sum_{\mu=2}^{p+2} 
	\psi^\mu_{-r} \widetilde\psi^\mu_{-r} 
+ \sum_{\mu=p+3}^{9} \psi^\mu_{-r} \widetilde\psi^\mu_{-r} 
\right] \right\} |Bp,\eta\rangle^{(0)}\nonumber \,.
\end{eqnarray}
Here $r$ is half-integer in the NS-NS sector, and integer in the R-R
sector. The ground state $|Bp,\eta\rangle^{(0)}$ is the tachyonic
ground state in the case of the NS-NS sector, and it is uniquely
determined by the condition (\ref{Dirichlet}) and (\ref{Neumann}) with
$r=0$ in the R-R sector. 

Given two coherent boundary states, we can calculate the closed string
tree amplitude that describes the exchange of closed string states
between the two boundary states. Because of world-sheet duality, this
amplitude can be reinterpreted as a one-loop open string vacuum
amplitude. The actual D-brane state is a certain linear combination of
these boundary states in different sectors and with different spin
structures \cite{BG1}; this linear combination is characterised by the
condition that it satisfies:
\begin{list}{(\roman{enumi})}{\usecounter{enumi}}
\item It is a {\em physical state} of the closed string theory,
{\it i.e.} it is GSO-invariant, and invariant under orbifold and
orientifold projections where appropriate.
\item The open string amplitude obtained by world-sheet duality
from the closed string exchange between any two D-branes
constitutes an {\em open string partition function}, {\it i.e.} 
it corresponds to a trace over a set of open string states
of the open string time-evolution operator.
\item The open strings that are introduced in this way have 
{\em consistent string field interactions} with the original closed
strings. 
\end{list}

In this talk we shall be mainly interested in {\em stable} D-branes;
this requires in particular that the spectrum of open strings that
begin and end on the same D-brane does not contain a tachyon. 

The conditions that we have outlined above are {\em intrinsic
consistency conditions} of the interacting string (field) theory; in
particular, they are more fundamental than spacetime supersymmetry,
and also apply in cases where spacetime supersymmetry is broken or
absent. 

\subsection{An example: Type IIA and IIB}

In order to demonstrate that these conditions capture the essence of
D-branes, let us analyse, as an example, the case of Type IIA/IIB
string theory. In the NS-NS sector, a GSO-invariant boundary state
exists for all $p$, 
\be
\label{NSNS}
|Bp\rangle_{NSNS} = \left( |Bp,+\rangle_{NSNS} - |Bp,-\rangle_{NSNS}
\right)\,,
\ee
but it does not describe a stable D-brane by itself since the open
string that begins and ends on $|Bp\rangle_{NSNS}$ consists of an
unprojected NS and R sector, and therefore contains a tachyon in its
spectrum. In fact (\ref{NSNS}) describes the unstable D$p$-brane for
$p$ odd (even) in Type IIA (IIB) that was described in Sen's talk; for
$p=9$, this coincides also with the unstable D9-brane of Type IIA that
was mentioned in Horava's talk. 

In order to obtain a stable D-brane, we have to add to (\ref{NSNS}) a
boundary state in the R-R sector; the only potentially GSO-invariant
state is of the form
\be
\label{RR}
|Bp\rangle_{RR} = \left( |Bp,+\rangle_{RR} + |Bp,-\rangle_{RR}
\right)\,,
\ee
and it is actually GSO-invariant if $p$ is even (odd) in the case of
Type IIA (IIB). If this is the case, we can find a suitable linear
combination\footnote{The sign in (\ref{Dp}) distinguishes a brane from
an anti-brane.}
\be
\label{Dp}
|Dp\rangle = {\cal N}_{NSNS} |Bp\rangle_{NSNS} \pm {\cal N}_{RR}
|Bp\rangle_{RR} \,,
\ee
whose corresponding open string spectrum consists of the GSO-projected
NS and R sector. Thus we have shown that the only stable D-branes 
in Type IIA and IIB are the familiar BPS D-branes.

\section{Non-BPS states in Heterotic -- Type II duality}

Many string theories contain states that are not BPS-saturated but are 
{\em stable} due to the fact that they are the lightest states of a
given charge \cite{Sen1}. Because of their stability, these states
must also be present in the dual theory. 

One particularly interesting example where both theories can be
analysed in detail is the duality between the heterotic string on
$T^4$ and Type IIA on K3 \cite{Witten1}. We shall in particular
consider the orbifold point of K3, where K3$=T^4/\Zop_2$, 
since one can then easily define boundary states in the IIA theory.
The sequence of dualities relating the two theories is given by   
\be
 \mbox{het}\;\; T^4 \stackrel{S}{\longrightarrow}
 \mbox{I}\;\; T^4 \stackrel{T^4}{\longrightarrow}
 \mbox{IIB}\;\; T^4/\bbbz_2' \stackrel{S}{\longrightarrow}
 \mbox{IIB}\;\; T^4/\bbbz_2'' \stackrel{T}{\longrightarrow}
 \mbox{IIA}\;\; T^4/\bbbz_2 \,,
\label{sequence}
\ee
where the various $\bbbz_2$ groups are 
\be
 \bbbz_2' = (1,\Omega\I_4) \quad
 \bbbz_2'' = (1,(-1)^{F_L}\I_4) \quad
 \bbbz_2 = (1,\I_4)\,.
\ee
Here $\I_4$ reflects all four compact directions, $\Omega$ reverses
world-sheet parity, and $F_L$ is the left-moving part of the spacetime
fermion number. 
It follows from (\ref{sequence}) that the parameters
of the two theories are related as 
\begin{eqnarray}
g_A & = & g_h^{-1} R_{h4} V_h^{1/2} \nonumber \vspace{0.2cm} \\
R_{Aj} & = & {2 \over R_{hj} } V_h^{1/2} \qquad \qquad \qquad
\mbox{for $j\ne 4$} \vspace{0.2cm} \label{duality} \\
R_{A4} & = & {1\over 2} V_h^{-1/2} R_{h4} \nonumber \,,
\end{eqnarray}
where $g_A, R_{Ai}$ and $g_h, R_{hi}$ are the coupling constant and
the radii of the IIA and the heterotic theory, respectively, and
$V_h= R_{h1} R_{h2} R_{h3} R_{h4}$.

At the orbifold point, the gauge group of the IIA theory is
$U(1)^{24}$; sixteen of these $U(1)$'s arise from the sixteen twisted
RR sectors that are associated to the fixed planes of the orbifold,
and the remaining eight $U(1)$'s come from the 1-form and the 3-form in
ten dimensions. In the dual heterotic theory (that we shall take to be
the $Spin(32)/\Zop_2$ theory), we also have to have an abelian gauge
group, and this requires that appropriate Wilson lines
are turned on 
\be
\begin{array}{rcl}
A^1 & = & \left(\left(\half\right)^8, 0^8 \right)   \\
A^2 & = & \left(\left(\half\right)^4,0^4,
           \left(\half\right)^4,0^4 \right) \\ 
A^3 & = & \left(\left(\half\right)^2,0^2,
           \left(\half\right)^2,0^2,\left(\half\right)^2,0^2,
           \left(\half\right)^2,0^2 \right) \\
A^4 & = &  \left(\half,0,\half,0,\half,0,\half,0,
                \half,0,\half,0,\half,0,\half,0 \right) \,.
\end{array}
\ee
These break the gauge group $SO(32)$ to $SO(2)^{16}\sim U(1)^{16}$. 
The actual form of the Wilson lines can be confirmed, 
{\it a posteriori}, by comparing the masses of the various BPS states.
Let us analyse some of them in turn.
\medskip

\noindent \underline{1. Bulk BPS D-particles}. The Type IIA orbifold
possesses a bulk D-particle, which corresponds to a Type IIA
D-particle together with its image under $\Zop_2$. This state carries
unit charge under the ten-dimensional R-R one-form $C_{RR}^{(1)}$, and
is described by 
\be
\label{D0}
\ket{D0;\epsilon} = \ket{U0}_{NSNS} + \epsilon \ket{U0}_{RR} \,,
\ee
where $\epsilon = \pm$ distinguishes the brane from the anti-brane,
and $\ket{U0}_{NSNS}$ and $\ket{U0}_{RR}$ are the GSO-invariant
combinations in (\ref{NSNS}) and (\ref{RR}), respectively. The
corresponding BPS state in the heterotic theory is a Kaluza-Klein
excitation ($N_L=1$) with momentum 
\be
\label{hetD0}
P_L = \left( 0^{16}; 0^3, \epsilon/R_{h4}\right)\,, \qquad \qquad
P_R = \left(0^3, \epsilon/R_{h4}\right)\,.
\ee
It can be shown (see \cite{BG3} for details) that the masses of the
two states agree.
\medskip

\noindent \underline{2. Fractional BPS D-particles}. The orbifold theory
also contains a `fractional' D-particle that is stuck at one of the
fixed planes \cite{DM}. In the blow up of the orbifold to a smooth K3,
this state corresponds to a D2-brane which wraps a supersymmetric
cycle \cite{Douglas}. In the orbifold limit the area of this cycle
vanishes, but the corresponding state is not massless, due to a
non-vanishing two-form flux $B$ through the cycle \cite{Aspinwall}. In
fact $B=1/2$, and the resulting state carries one unit of
twisted-sector charge coming from the membrane itself, and one half
unit of D-particle charge coming from the D2-brane world-volume action
term $\int d^3\sigma \, C_{RR}^{(1)}\wedge (F^{(2)}+B^{(2)})$. 
The corresponding boundary state is of the form  
\be
\fl
\label{D0frac}
\ket{D0_f;\epsilon_1,\epsilon_2} = {1 \over 2}\left[
\Big(\ket{U0}_{NSNS} + \epsilon_1\ket{U0}_{RR}\Big) 
+ \epsilon_2\Big(\ket{T0}_{NSNS} +\epsilon_1\ket{T0}_{RR}\Big)\right]
\,,  
\ee 
where $\ket{U0}_{NSNS}$ and $\ket{U0}_{RR}$ are the same states that
appeared in (\ref{D0}), and $\ket{T0}_{NSNS}$ and $\ket{T0}_{RR}$ lie
in the twisted NS-NS and twisted R-R sectors, respectively. Here
$\epsilon_1=\pm 1$ and $\epsilon_1\epsilon_2=\pm 1$ determine the sign
of the bulk and the twisted charges of the state, respectively. 
As there are $16$ fixed planes, there are $64$ such states
altogether. 

The corresponding states in the heterotic theory carry 
momentum
\be
\fl
\label{hetfrac}
P_L = \left(\epsilon_1\epsilon_2(0^{2n},1,\pm 1, 0^{14-2n}); 0^3,
\epsilon_1/(2 R_{h4}) \right) \,, \qquad \qquad
P_R= \left(0^3,\epsilon_1/(2 R_{h4}) \right) .
\ee
The $16$ vectors $(0^{2n},1,\pm 1, 0^{14-2n})$, with
$n=1,\ldots, 8$, are in one-to-one correspondence with the $16$
fixed planes in the IIA orbifold. For each such vector there are 
four different heterotic states. These states carry half the charge of
the states in (\ref{hetD0}) with respect to the $U(1)$ that is
associated to the KK momentum, and their mass is therefore also
half of that of the states in (\ref{hetD0}). This is mirrored in the
IIA theory, where the coefficients of the boundary state (\ref{D0frac})
in the untwisted NS-NS and untwisted R-R sector (from which the mass and
charge can be read off) are half of those in
(\ref{D0}). 

\subsection{Non-BPS states}

The simplest stable non-BPS state in the heterotic theory has momentum
\be
\label{hetnon}
P_L = \left(0^{2n},2,0^{15-2n}; 0^4 \right) \,, \qquad \qquad
P_R= \left(0^{4}\right) \,.
\ee
Level matching requires that $N_R-c_R=1$, and the state is therefore
not BPS \cite{dh}. The mass of this state is $M_h=2 \sqrt{2}$,
\footnote{In our conventions, $\alpha'_h=1/2$.} 
and it is charged under precisely two of the $16$ $U(1)$'s that are
associated with the $16$ fixed planes in the dual Type IIA theory, and
uncharged with respect to any of the other $U(1)$'s. More
generally, for each pair of fixed planes, there exist four
non-BPS states that carry charge $\pm 1$ with respect to the two
corresponding $U(1)$'s, but are uncharged with respect to any other
$U(1)$. 

The above state carries the same charges as two BPS states with
charges 
\be
\label{hetBPS}
\fl
\begin{array}{ll}
{\displaystyle P^{(1)}_L = \left( 0^{2n},1, 1, 0^{14-2n}; 0^3, 
1/(2 R_{h4}) \right) \,,} \qquad \qquad & 
{\displaystyle
P^{(1)}_R= \left(0^3,1/(2 R_{h4}) \right) \,,} \\
{\displaystyle P^{(2)}_L = \left( 0^{2n},1,-1, 0^{14-2n}; 0^3, 
-1/(2 R_{h4})\right) \,,  \qquad \qquad} & 
{\displaystyle
P^{(2)}_R= \left(0^3,-1/(2 R_{h4}) \right) \,.} \\
\end{array}
\ee
Each of the two BPS states in (\ref{hetBPS}) has mass 
$1/(2 R_{h4})$, and the decay of (\ref{hetnon}) into (\ref{hetBPS})
is energetically forbidden provided that 
$R_{h4} < 1/(2 \sqrt{2})$. Similarly, we can analyse the other
decay channels, and we find that (\ref{hetnon}) is stable provided
that 
\be
\label{hstability}
R_{hj} < {1 \over 2 \sqrt{2}} \qquad j=1,2,3,4\,.
\ee
As we have seen above, the two BPS states in (\ref{hetBPS})
correspond, in the dual Type IIA theory, to two fractional BPS
D-particles that are localised at different fixed planes, and that have
{\em opposite} bulk charge. The non-BPS state (\ref{hetnon}) therefore
corresponds to a non-BPS D-string that stretches between a pair of
fixed planes; in terms of boundary states this non-BPS state can be
described as\footnote{This state has also been independently
constructed by Sen \cite{Sen5}; compare also Sen's contribution to
these proceedings.} 
\be
\fl
\label{D1p}
\ket{D1_{nonbps};\theta,\epsilon} =
  {1\over\sqrt{2}}\left[\ket{U1;\theta}_{NSNS} +
  {\epsilon\over\sqrt{2}} 
  \Bigl( \ket{T1;1}_{RR} + e^{i\theta} \ket{T2;2}_{RR} \Bigr)\right]
  \,.   
\ee
Here $\theta=0,\pi$ describes the Wilson line on the D-string, and
$\ket{U1;\theta}_{NSNS}$ is defined by 
\be 
\ket{U1;\theta}_{NSNS} = \sum_{w} e^{i\theta w} \ket{U1;w}_{NSNS} \,, 
\ee 
where $w$ denotes the winding number along the direction of the
D-string. The two states in the twisted R-R sector $\ket{T1;1}_{RR}$
and $\ket{T1;2}_{RR}$ are localised at either end of the D-string (so
that the D-string stretches between two fixed planes). Using standard
techniques \cite{BG1,Sen2}, one can easily check that each of the
boundary state components is invariant under the GSO and orbifold
projections. Furthermore, the spectrum of open strings beginning and
ending on the same D-string can be obtained, as usual, from the
cylinder amplitude of the above boundary state using world-sheet
duality; for a suitable normalisation of the different components,
this leads to \be [NS - R ] \; {1 \over 4} \left( 1+ (-1)^F \I_4
\right) \left( 1 +(-1)^F \I'_4 \right) \,,
\label{open_D1}
\ee
where $\I'_4$ is the same as $\I_4$, except that it acts on
$x^4$ as $x^4\rightarrow 2\pi R_{A4} - x^4$. The boundary state
(\ref{D1p}) therefore satisfies the consistency condition (ii). 
Sen has also argued \cite{Sen4,SenRev} that (iii) is satisfied.

For each pair of fixed planes there are four such D-strings, that
carry charge $\pm 1$ (depending on the four choices of $\theta=0,\pi$
and $\epsilon=\pm 1$) with respect to the two twisted sector $U(1)$'s 
associated to the two fixed planes. These charges are of the same 
magnitude as those of the fractional D-particles, since the ground
state of $\ket{T1}_{RR}$ is the same as that of $\ket{T0}_{RR}$ in
(\ref{D0frac}). Furthermore, it follows from (\ref{open_D1}) that the
D-strings have sixteen (rather than eight) fermionic zero modes, and
therefore transform in long multiplets of the $D=6$, ${\cal N}=(1,1)$
supersymmetry algebra. These states therefore have exactly the correct
properties to correspond to the above non-BPS states of the heterotic
theory.   

The D-string that corresponds to the state (\ref{hetnon}) stretches
along the $x^4$ coordinate in Type IIA, and it is stable provided that
\be
\label{Astability}
R_{A4} < \sqrt{2} \qquad \mbox{and} \qquad
R_{Aj} > {1 \over \sqrt{2}}\,, \quad j=1,2,3 \,.
\ee
Indeed, if the former inequality is violated, the open string spectrum
in (\ref{open_D1}) contains a tachyon of unit KK momentum in the $x^4$
direction; if the latter inequality is violated, the spectrum contains
a tachyon of unit winding in the $x^1, x^2$, or $x^3$ direction.
These inequalities can also be determined by comparing the mass
of the D-string with that of a pair of fractional BPS D-particles
in the former case, and a pair of fractional BPS D2-branes in the
latter. 

The domains of stability (\ref{hstability}) and (\ref{Astability}) are
{\em qualitatively} related by the duality map (\ref{duality}).
Since the states in question are not BPS, the masses are not protected
from quantum corrections, and one should not expect that the regimes of
stability match exactly.

\subsection{T-duality}

T-duality relates the non-BPS D-string (\ref{D1p}) of the Type IIA
orbifold to a non-BPS D-particle of Type IIB on $T^4/(-1)^{F_L} \I_4$
\cite{BG2,Sen2}. (Under S-duality, this orbifold is related to the
orientifold of Type IIB on $T^4$ by $\Omega \I_4$ \cite{Sen1}, and the
non-BPS D-particle corresponds to the first excited state of the string
that stretches between the D5-brane and its mirror \cite{BG2,Sen2}.) 
As we have seen above, the non-BPS D-string
can decay into two BPS D-particles that sit at opposite ends of the
D-string and carry opposite bulk charge. Under T-duality,
this configuration corresponds to a BPS D-string and an
anti-D-string that carry a relative Wilson line. The above analysis
therefore gives support for the claim that the non-BPS D-particle can
indeed be described in terms of a brane-anti-brane pair \cite{Sen2}.

\subsection{Other non-BPS states}

The heterotic string theory also contains stable non-BPS states that
do not correspond to D-branes in the IIA orbifold. 
The simplest examples are the
states that transform in the spinor representation of $SO(32)$. In
$D=10$ these states have been identified with a $\bbbz_2$-valued 
non-BPS D-particle in the dual type I string
\cite{Sen3,Sen4,Witten2}. The sequence of duality transformations that
relate the heterotic string on $T^4$ to Type IIA on K3 (\ref{sequence}) 
suggests that
in six dimensions these states correspond to a non-BPS (non-Dirichlet)
4-brane, which may be understood as a bound state of an NS5-brane and
an anti-NS5-brane. 
\smallskip  

One can also compare non-BPS states that are not necessarily
stable. For example, the heterotic theory contains states that are
charged with respect to a {\em single} $U(1)$ associated with one
fixed plane in the dual IIA orbifold, and are uncharged with 
respect to any other $U(1)$; one such state is of the form
\be
\label{hetnon1}
P_L = \left(2,2,0^{16}; 0^4 \right) \,, \qquad \qquad
P_R= \left(0^{4}\right) \,.
\ee
The mass of this state is $M_h = 2 \sqrt{6}$, and it is actually
unstable in the heterotic theory\footnote{The following discussion
corrects the discussion of \cite{BG3}, where winding states were not
considered.}. Indeed, (\ref{hetnon1}) can decay into four BPS states of
the form   
\be\label{decay1}
\fl
\begin{array}{rclrcl}
P_L^{(1)} & = & (1,0,1,0^{13};0,0,1/(2 R_{h3}),0) &
P_R^{(1)} & = & (0,0,1/(2 R_{h3}),0) \\
P_L^{(2)} & = & (1,0,-1,0^{13};0,0,-1/(2 R_{h3}),0)  &
P_R^{(1)} & = & (0,0,-1/(2 R_{h3}),0) \\
P_L^{(3)} & = & (0,1,0,1,0^{12};0,0,1/(2 R_{h3}),0) &
P_R^{(3)} & = & (0,0,1/(2 R_{h3}),0) \\
P_L^{(4)} & = & (0,1,0,-1,0^{12};0,0,-1/(2 R_{h3}),0) \qquad &
P_R^{(4)} & = & (0,0,-1/(2 R_{h3}),0) \,.
\end{array}
\ee
The mass of each of these states is $1/(2R_{h3})$, and this decay
process is energetically forbidden provided that 
\be \label{bound}
R_{h3} < {1 \over \sqrt{6}}\,.
\ee
However, (\ref{hetnon1}) can also decay into four winding BPS states
whose left-moving (shifted) momenta are given by
\be\label{decay2}
\fl
\begin{array}{rcl}
P_L^{(1)} & = &
({1\over 2},{1\over 2},0,0,-{1\over 2},{1\over 2},0,0,
{1\over 2},-{1\over 2},0,0,{1\over 2},{1\over 2},0,0;0,0,R_{h3},0) \\
P_L^{(2)} & = &
({1\over 2},{1\over 2},0,0,{1\over 2},-{1\over 2},0,0,
-{1\over 2},{1\over 2},0,0,{1\over 2},{1\over 2},0,0;0,0,-R_{h3},0)  
\\
P_L^{(3)} & = &
({1\over 2},{1\over 2},0,0,-{1\over 2},{1\over 2},0,0,
-{1\over 2},{1\over 2},0,0,-{1\over 2},-{1\over 2},0,0;0,0,R_{h3},0) 
\\
P_L^{(4)} & = &
({1\over 2},{1\over 2},0,0,{1\over 2},-{1\over 2},0,0,
{1\over 2},-{1\over 2},0,0,-{1\over 2},-{1\over 2},0,0;
0,0,-R_{h3},0)\,,
\end{array}
\ee
where the corresponding right-moving momenta equal again the last
four entries. Each of these has mass $2 R_{h3}$, so this decay process
is forbidden when 
\be\label{bound2}
R_{h3} > {\sqrt{6}\over 4} = {3\over 2\sqrt{6}}\,.
\ee
It therefore follows that the non-BPS state is unstable for all
values of $R_{h3}$.

This is actually mirrored in the dual Type IIA theory, where
(\ref{hetnon1}) corresponds to a state that has the same charges as
two fractional BPS D-particles that are located at the same fixed plane
but carry opposite bulk charge. As was shown in \cite{BG3}, the theory
does not contain a stable D-brane with these
charges. Furthermore, the possible bound state of the two D-particles
does not exist, since a careful analysis of the potential demonstrates
that the interaction is always repulsive \cite{GabSen}.

\section{Conclusions}

We have demonstrated how non-BPS D-branes can be used to
probe string dualities beyond the BPS spectrum. For the case of the
heterotic string on $T^4$ and Type IIA on K3 that we have analysed in
detail, both sides are quantitatively under control, and one can
compare the stability of the different non-BPS states. We have found
that the domains of stability are qualitatively related by the duality
map. It would be interesting to analyse these non-BPS states at more
generic points in the moduli space of K3; first steps in this
direction have recently been taken in \cite{Sen5,MajSen}.

The techniques that we have described here should shed further light
on the dualities that have been proposed for some non-supersymmetric
theories, in particular for the models considered  by Kachru, Kumar \&
Silverstein \cite{KKS} and Harvey \cite{Harvey} (see also
\cite{Koers}).

\section*{Acknowledgements}

We thank Ashoke Sen for useful conversations.

\noindent O.B. is supported in part by the DOE under grant
no. DE-FG03-92-ER 40701.  M.R.G. is supported by a College Lectureship
of Fitzwilliam College, Cambridge.

\end{document}